\documentclass[conference]{IEEEtran}
\IEEEoverridecommandlockouts

\usepackage{cite}
\usepackage{amsmath,amssymb,amsfonts}
\usepackage{algorithmic}
\usepackage{graphicx}
\usepackage{textcomp}
\usepackage{xcolor}
\usepackage{multicol}
\usepackage{multirow}
\usepackage{subcaption}
\usepackage{amsmath}

\def\BibTeX{{\rm B\kern-.05em{\sc i\kern-.025em b}\kern-.08em
    T\kern-.1667em\lower.7ex\hbox{E}\kern-.125emX}}
\begin{document}

\title{DMT: Demographic Conditioning, Morphology-Enhanced Transformer for Cuffless Blood Pressure Estimation from PPG Signals}

\author{\IEEEauthorblockN{\small Yidan Shen}
\IEEEauthorblockA{\scriptsize\textit{Dept. of Electrical and Computer Engineering} \\
\textit{University of Houston}\\
Houston, USA \\
yshen20@uh.edu}
\and
\IEEEauthorblockN{\small Neville Mathew}
\IEEEauthorblockA{\scriptsize\textit{Dept. of Engineering Technology} \\
\textit{University of Houston}\\
Houston, USA \\
namathew3@uh.edu}

\and
\IEEEauthorblockN{\small Maham Rahimi}
\IEEEauthorblockA{\scriptsize\textit{Dept. of Cardiovascular Surgery} \\
\textit{Houston Methodist Hospital}\\
Houston, USA \\
mrahimi@houstonmethodist.org}
\and

\IEEEauthorblockN{\small Deependra Dhakal}
\IEEEauthorblockA{\scriptsize\textit{Dept. of Engineering Technology} \\
\textit{University of Houston}\\
Houston, USA \\
ddhakal@cougarnet.uh.edu}
\and

\IEEEauthorblockN{\small George Zouridakis}
\IEEEauthorblockA{\scriptsize\textit{Dept. of Engineering Technology}\\
\textit{University of Houston}\\
Houston, USA \\
zouridakis@uh.edu}
\and
\IEEEauthorblockN{\small Xin Fu}
\IEEEauthorblockA{\scriptsize\textit{Dept. of Electrical and Computer Engineering} \\
\textit{University of Houston}\\
Houston, USA \\
xfu8@central.uh.edu}
\and

\IEEEauthorblockN{\small Renjie Hu$^{*}$ \thanks{$*$ Corresponding author: Renjie Hu}}
\IEEEauthorblockA{\scriptsize\textit{Dept. of Information Science Technology} \\
\textit{University of Houston}\\
Houston, USA \\
rhu7@uh.edu}
}

\maketitle

\begin{abstract}
Blood pressure (BP) is a key marker for cardiovascular risk assessment and therapeutic decision-making, and Photoplethysmography (PPG) enables low-cost, wearable-friendly cuffless BP estimation. However, even with recent progress, many PPG-based models are trained with BP regression alone and may rely on amplitude-dominated shortcuts. In addition, demographic covariates that systematically modulate vascular compliance are often incorporated only via late fusion, limiting subject-specific representation learning.
We propose a Transformer-based network for cuffless BP estimation from PPG signal, leveraging self-attention to capture long-range dependencies across multiple cardiac cycles. To account for subject-specific vascular differences, the model is conditioned on demographics via FiLM-style feature modulation applied through the attention and feed-forward sublayers of Transformer blocks. In addition, we add an auxiliary morphology head to guide the model to attend to BP-relevant waveform morphology associated with arterial stiffness and wave reflection.
Under calibration-based evaluation protocols on the large-scale PulseDB dataset, the proposed method achieves MAE of 4.56 mmHg for systolic BP and 2.62 mmHg for diastolic BP, reducing errors by 47\% and 50\% compared with prior demographic-enhanced PPG baselines. The resulting lightweight, single-sensor model supports scalable and clinically grounded cuffless BP estimation in calibration-enabled deployment settings.
\end{abstract}


\section{Introduction}
Blood pressure (BP) is a key physiological parameter that underpins cardiovascular risk assessment and therapeutic decision-making. Both systolic blood pressure (SBP) and diastolic blood pressure (DBP) are routinely measured in clinical practice and are fundamental to the diagnosis and management of hypertension, a condition strongly associated with myocardial infarction, stroke, and chronic kidney disease \cite{oparil2018hypertension, magder2018meaning}. Cuff-based sphygmomanometers remain the clinical gold standard, but their use is limited to intermittent, point-in-time readings, and is often inconvenient for patients, leading to poor adherence in daily-life monitoring \cite{tale2021sphygmomanometers}. These limitations highlight the need for cuffless and continuous solutions.

Photoplethysmography (PPG) has emerged as a promising modality for cuffless BP estimation. By measuring peripheral blood volume changes optically, PPG offers a low-cost, noninvasive, and wearable-friendly option for cardiovascular monitoring. Its waveform reflects both hemodynamic parameters and arterial compliance, carrying rich information about vascular tone, stiffness, and wave reflection \cite{allen2007ppg, elgendi2012analysis}. Unlike multi-sensor methods that require electrocardiography (ECG) measurements, PPG-based single-sensor pathways reduce hardware complexity and improve the feasibility of integration into consumer devices \cite{ELHAJJ2021102301}.

Machine learning has been widely applied across healthcare \cite{qin2025explainablecounterfactualreasoningdepression, 11254303}. Early deep learning methods enable end-to-end BP prediction from raw PPG, commonly using 1D convolutional backbones \cite{sadrawi, moulaeifard2025generalizabledeeplearningphotoplethysmographybased}. While effective at extracting local patterns, convolutional models are biased toward short-range correlations and can under-utilize cross-cycle dependencies that are relevant to reflection-related structures and slow modulations. To improve temporal scalability, sequence models capable of efficient long-context modeling have also been adopted in PPG-based BP estimation \cite{gu2022efficientlymodelinglongsequences, moulaeifard2025generalizabledeeplearningphotoplethysmographybased}. Recent studies increasingly incorporate global temporal modeling—via Transformer-based architectures or hybrid CNN–Transformer designs—to better capture long-range dependencies beyond purely convolutional receptive fields \cite{charlton2022vascularage, tian2025pctn, arjomand2024transforhythm}. In addition, several works enhance BP estimation by fusing demographic information (e.g., age, sex, BMI) through late-fusion to account for inter-subject variability \cite{liu2024bigru_attention, ajmal2021obesityppg}.

\begin{figure}[t]
    \centering
    \includegraphics[width=0.48\textwidth]{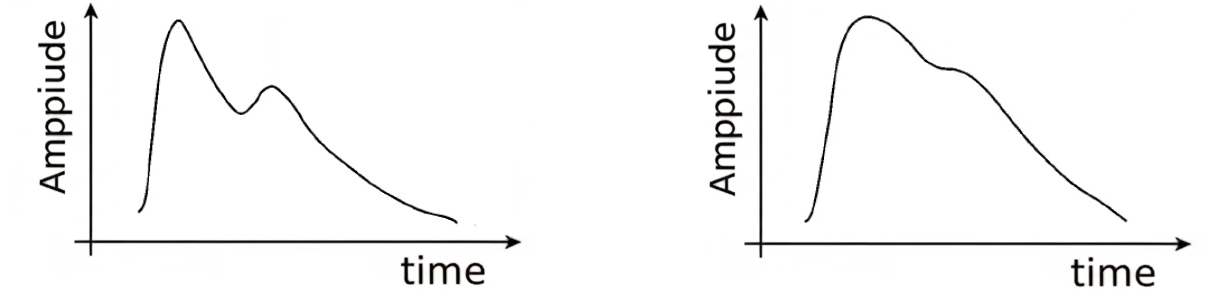}
    \caption{Representative examples of PPG waveform morphology in normotensive (left) and hypertensive (right) subjects.}
    \label{fig:wave} 
\end{figure}

Physiological and clinical studies have shown that BP, vascular properties, and PPG waveform morphology are tightly coupled. As illustrated in Fig.~\ref{fig:wave}, hypertensive PPG waveforms typically exhibit a flattened or plateaued systolic apex, in contrast to the sharper and more peaked systolic upstroke observed in normotensive subjects \cite{shimizu2008aixreview, katsuda2013aix, allen2007ppg, liang2018ppgaging}. These morphology-level differences reflect increased arterial stiffness and altered wave reflection, and encode information about both instantaneous BP and long-term vascular condition.
Demographic factors further modulate these vascular properties and their expression in PPG signals. Age, sex, and BMI influence arterial compliance, wave reflection characteristics, and signal attenuation, leading to systematic, subject-dependent variations in PPG morphology \cite{allen2007ppg, liang2018ppgaging, elgendi2012analysis, rodriguez2022obeseoptics, avolio1986gender}. Collectively, prior evidence indicates that demographic information should act as contextual priors that shape how PPG waveform features are extracted and interpreted in a model. 

However, two gaps remain. First, demographic covariates are often appended via late-fusion to the final regressor and thus do not modulate intermediate feature extraction, limiting the model’s ability to internalize subject-specific vascular priors. Second, many end-to-end models optimize BP regression alone, without explicit morphology-oriented supervision, which can encourage amplitude-dominated shortcuts and reduce robustness across subjects.

To address these limitations, we propose a demographic conditioning, morphology-enhanced Transformer for cuffless BP estimation from PPG. We first tokenize long, densely sampled PPG segments using a 1D convolutional patch-embedding with learnable positional embedding, which enables efficient modeling of long-range temporal dependencies across multiple cardiac cycles. To incorporate subject context, we embed age, sex, and BMI, then use a FiLM generator to produce layer and feature-wise modulation parameters that are injected into both the attention and feed-forward sub-layers of every Transformer block. This design keeps the core attention mechanism shared while allowing demographic priors to modulate feature extraction throughout the network. Finally, we introduce an auxiliary head to predict a hypertension-related morphology label from the global representation and optimize BP regression and morphology classification jointly with two learnable loss weights, encouraging morphology-sensitive representations beyond amplitude cues. Experiments on a large scale PulseDB dataset \cite{pulsedb_paper} show that our method achieves mean absolute errors (MAE) of 4.56 mmHg for SBP and 2.62 mmHg for DBP under calibration-based evaluation protocols, representing reductions of 47\% and 50\%, respectively, compared with prior demographic-enhanced models. Moreover, the proposed approach yields a lightweight, single-sensor model well suited for wearable deployment.

To conclude, our main contributions are:

\begin{itemize}
    \item We propose a Transformer-based architecture for cuffless blood pressure estimation using single PPG sensor, enabling continuous and wearable-friendly BP monitoring.

    \item We utilize a PPG tokenization framework based on a 1D convolutional patch-embedding with positional embedding, which efficiently captures long-range temporal dependencies across cardiac cycles.

    \item We introduce a FiLM-based demographic conditioning mechanism that embeds age, sex, and BMI to modulate both attention and feed-forward sublayers in every Transformer block, enabling subject-specific vascular priors to shape feature extraction throughout the network.

    \item We incorporate morphology-oriented supervision by jointly optimizing BP regression and PPG morphology classification with learnable loss weights.

    \item Extensive experiments demonstrate that the proposed method significantly outperforms prior models.
\end{itemize}

\section{Related Works}
There are two categories that dominate cuffless BP estimation: (i) PPG with auxiliary physiology signals, which typically improves stability but requires multi-sensor inputs; and (ii) PPG-only models, which maximize portability but are sensitive to motion, perfusion, and device variability.

\subsection{PPG with auxiliary physiology signals}
Early systems fused PPG with timing cues such as PAT, PTT or time-to-peak (TTP) \cite{pat_paper, pwv_paper}, often with linear or polynomial regression. These pipelines work in controlled settings but depend on reliable R-peak detection and precise sensor application; even a small calibration drift or a missing PPG signal can degrade accuracy and limit real-world prediction estimates.
Deep-BP \cite{yan2019novel} introduced a CNN with hard parameter-sharing multi-task learning to jointly estimate SBP and DBP from synchronized ECG+PPG signals, using mean-filter denoising and classic MTL sharing \cite{caruna1993multitask}. \cite{huang2022mlp} adapted MLP-BP for gMLP/LSTM-Mixer-style token mixing to multi-channel ECG+PPG, aiming to replace hand-crafted features.

While these approaches have advanced BP estimation, they remain constrained by multi-sensor dependency and model ability which limits scalability and real-world wearability.

\subsection{PPG-only}
PPG-only methods maximize portability and are easiest to integrate into wearable devices.
Early machine learning approaches mapped features extracted from PPG and its derivatives to BP values using regression models such as support vector machines, random forests, or boosting ensembles \cite{chu2023noninvasive,nie2024reviewdeeplearningmethods}.
Sadrawi et al. formulate genetic deep convolutional autoencoders (LeNet and U-Net-based DCAE ensemble models) to reconstruct ABP from single-channel PPG and derive BP from the recovered waveform \cite{sadrawi}. The idea stabilizes learning on short segments, but the study uses a small cohort with manual filtering of atypical waveforms—conditions that under-represent real-world noise and motion. 
Moulaeifard et al. retrofit 2D convolution networks for 1-D PPG and emphasize standardized reporting (AAMI/BHS) with external reviews \cite{moulaeifard2025generalizabledeeplearningphotoplethysmographybased}. 
TCN-style architectures with dilated causal convolutions and attention modules have been proposed to better capture longer temporal dependencies \cite{dai2024tcn_attention}.
Multi-task designs also leverage PPG and its derivatives (VPG/APG) and jointly optimize correlated BP targets to enhance feature extraction and fusion \cite{xiao2024mdlqmtlnet}.
Hybrid CNN-Transformer frameworks combine local convolution with global attention for BP estimation \cite{tian2025pctn}, and Transformer-based pipelines have also been evaluated on ICU waveform corpora for PPG-only BP estimation \cite{arjomand2024transforhythm}.
Prior studies have also incorporated demographic with PPG signals via late-fusion strategies on CNN or sequential models in healthy adult cohorts. \cite{biosignals26} Similarly, Liu et al. incorporate demographic covariates for personalization and trend estimation, by concatenating demographics with the learned PPG representation near the output layer \cite{liu2024bigru_attention}.

Despite these advances, most PPG-only approaches still (i) incorporate demographics via late fusion rather than feature-level, and (ii) optimize BP regression objectives alone without explicit morphology-oriented auxiliary supervision tied to hypertensive waveform phenotypes.

\section{Methodology}

\begin{figure}[t]
    \centering
    \includegraphics[width=0.51\textwidth]{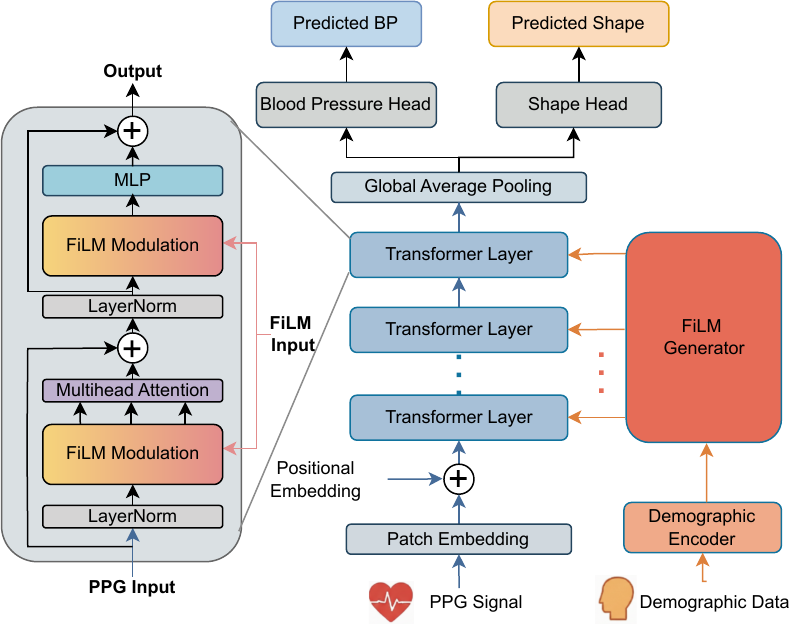}
    \caption{Network architecture. We use two separate networks, one dedicated to SBP and the other to DBP.}
    \label{fig:arch} 
\end{figure}

To jointly capture long-range PPG dynamics, subject-level demographic priors, and physiologically meaningful waveform morphology, we design a Transformer architecture with FiLM-based demographic conditioning.
The architecture has three components: (i) a patch embedding stem that converts raw PPG waveforms into a compact token sequence via 1D convolution with learnable positional embeddings; (ii) a demographic pathway that encodes static covariates (age, sex, BMI) and generates feature-wise linear modulation (FiLM) parameters for every Transformer layer, thereby injecting subject-specific priors into both attention and feed-forward layers; and (iii) dual heads for BP regression and morphology supervision that encourage the backbone to encode physiologically meaningful pulse-shape patterns rather than relying on amplitude-only cues. The architecture is shown as Fig.~\ref{fig:arch}.

\subsection{Patch Embedding}

Raw PPG segments contain a large number of densely sampled data points and are affected by noise.
Instead of feeding the full sequence directly into self-attention, we discretize it into non-overlapping patches of width $P$ using a 1D convolution layer, which compresses long PPG sequences into $T$ tokens while preserving fine-grained morphology within each patch. 
This reduces noise sensitivity, lowers the quadratic attention cost from $O(L^2)$ to $O(T^2)$ ($T<<L$), and makes the representation more robust to local distortions.

Because PPG is a time-series signal, we add a learnable positional embedding $\mathbf{E}_{\text{pos}} \in \mathbb{R}^{1 \times T \times C}$ to retain ordering information that is critical for physiological relationships:
\begin{equation}
\mathbf{X}_0 \;=\; \mathbf{Z}_0 + \mathbf{E}_{\text{pos}}.
\end{equation}

\subsection{Demographic Encoder and FiLM Generator}

Static covariates $s \in \mathbb{R}^3$ are first standardized/binarized during pre-processing.
We encode $s$ through a multi-layer perceptron (MLP) to obtain a demographic embedding $\mathbf{d} \in \mathbb{R}^C$.

To inject demographic priors into all Transformer layers, we use a FiLM generator that expands $\mathbf{d}$ into layer-wise, feature-wise modulation parameters. 
For a depth-$D$ encoder, the generator outputs:
\begin{equation}
\mathbf{F} \;=\; g_{\text{FiLM}}(f_{\text{demo}}(s)) 
\;\in\; \mathbb{R}^{D \times 4 \times C},
\end{equation}
The tensor $\mathbf{F}$ is then reshaped into per-layer parameter tuples:
\begin{equation}
\bigl(\gamma^{(1)}_\ell, \beta^{(1)}_\ell, \gamma^{(2)}_\ell, \beta^{(2)}_\ell\bigr) \in \mathbb{R}^{C}, 
\quad \ell = 1,\dots,D.
\end{equation}
Here $(\gamma^{(1)}_\ell, \beta^{(1)}_\ell)$ modulate the input to the attention sublayer, and $(\gamma^{(2)}_\ell, \beta^{(2)}_\ell)$ modulate the input to the feed-forward MLP in layer $\ell$. 
This design allows demographic priors to shape how each layer processes temporal information, without changing the underlying attention operator, increasing its parameter count or input length.

\subsection{FiLM-Conditioned Transformer Encoder}

Each encoder block is a Transformer with demographic FiLM applied on both residual branches. 
Given the token sequence $\mathbf{X}_{\ell-1} \in \mathbb{R}^{B \times T \times C}$ at layer $\ell$, we first apply layer normalization followed by FiLM modulation before self-attention:
\begin{align}
\hat{\mathbf{X}}^{(1)}_\ell 
&= \mathrm{LN}(\mathbf{X}_{\ell-1}) \odot \bigl(1 + \gamma^{(1)}_\ell\bigr) + \beta^{(1)}_\ell,
\end{align}
where $\gamma^{(1)}_\ell, \beta^{(1)}_\ell \in \mathbb{R}^{1 \times 1 \times C}$ are broadcast along the temporal dimension, and $\odot$ denotes element-wise multiplication. 
Multi-head self-attention is then computed as
\begin{equation}
\mathbf{H}_\ell = \mathrm{MHA}(\hat{\mathbf{X}}^{(1)}_\ell W_Q^{(\ell)},\hat{\mathbf{X}}^{(1)}_\ell W_K^{(\ell)},\hat{\mathbf{X}}^{(1)}_\ell W_V^{(\ell)}),
\end{equation}

and the first residual update is
\begin{equation}
\mathbf{X}'_\ell = \mathbf{X}_{\ell-1} + \mathbf{H}_\ell.
\end{equation}

We then apply a second FiLM-modulated pre-norm before the feed-forward network:
\begin{align}
\hat{\mathbf{X}}^{(2)}_\ell 
&= \mathrm{LN}(\mathbf{X}'_\ell) \odot \bigl(1 + \gamma^{(2)}_\ell\bigr) + \beta^{(2)}_\ell, \\
\mathbf{X}_\ell 
&= \mathbf{X}'_\ell + \mathrm{MLP}(\hat{\mathbf{X}}^{(2)}_\ell).
\end{align}

By modulating both the attention and MLP inputs, FiLM acts as a demographic-dependent ``gate'' that can re-weight, amplify, or attenuate specific feature channels in a subject-specific manner, while keeping the attention kernel itself shared across the population. 
Intuitively, this allows the network to emphasize faster, high-frequency systolic components for younger subjects, or slower reflection waves and broadened systolic peaks for older or higher-BMI subjects, without overfitting a separate model per subgroup.
After $D$ such layers, we obtain the final token sequence $\mathbf{X}_D$.

\subsection{Global Readout and Prediction Heads}

The token sequence $\mathbf{X}_D$ is normalized along feature dimension and aggregated over temporal axis to form a compact representation $\mathbf{g}$. Then we instantiate two prediction heads with disentangled objectives: 
a blood-pressure regression head that maps $\mathbf{g}$ to a single BP component $y\in\{\text{SBP}, \text{DBP}\}$; we train two identical networks independently for SBP and DBP to avoid cross-component interference under the one-task-per-model setting; and an auxiliary morphology head that classifies hypertension-related pulse-shape phenotypes, encouraging the backbone to encode systolic upstroke sharpness, dicrotic notch prominence, and late-wave reflection patterns tied to vascular stiffness rather than relying on amplitude-only cues. 
This multi-task design injects a physiologically meaningful inductive bias and improves robustness via representation sharing.

\subsection{Loss Function}

We formulate BP estimation as a multi-task learning problem that jointly optimizes both continuous BP regression and pulse morphology classification. Rather than manually assigning fixed loss weights, we adopt a learnable uncertainty-based weighting strategy to automatically balance the two tasks. Specifically, the overall objective is defined as:

\begin{equation}
\mathcal{L}
= \frac{1}{\sigma_{\text{bp}}}\,\mathcal{L}_{\text{bp}} + \log \sigma_{\text{bp}}
+ \frac{1}{\sigma_{\text{shape}}}\,\mathcal{L}_{\text{shape}} + \log \sigma_{\text{shape}},
\end{equation}

where $\mathcal{L}_{\text{bp}}$ denotes the MAE for systolic or diastolic BP regression, $\mathcal{L}_{\text{shape}}$ is the cross-entropy loss for morphology classification, and $\sigma_{\text{bp}}$ and $\sigma_{\text{shape}}$ are positive, learnable scalars.
This formulation can be interpreted as modeling task-dependent homoscedastic uncertainty.
Tasks with higher intrinsic noise are assigned larger $\sigma$, which down-weights their contribution to the total loss, while still penalizing excessive uncertainty through the logarithmic regularization terms.
By adaptively weighting BP regression and morphology supervision, the multi-task objective leverages morphology prediction as an auxiliary regularizer that improves BP estimation accuracy and generalization.

\section{Experiment}

\begin{table*}[htbp]
\caption{Model Performance Comparison. (P: PPG only; P,D: PPG + Demographic information)}
\centering
\begin{tabular}{|l|c|c|c|c|c|c|c|c|c|c|}
\hline
\multirow{2}{*}{\textbf{Model}} & \multirow{2}{*}{\textbf{input}} 
& \multicolumn{3}{c|}{\textbf{Cal-base (SBP/DBP)}} 
& \multicolumn{3}{c|}{\textbf{Cal-free (SBP/DBP)}} 
& \multicolumn{3}{c|}{\textbf{AAMI (SBP/DBP)}} \\
\cline{3-11}
& & \textbf{MAE} & \textbf{std} & \textbf{R²} & \textbf{MAE} & \textbf{std} & \textbf{R²} & \textbf{MAE} & \textbf{std} & \textbf{R²}\\
\hline

LeNet & P & 13.26/8.09 & 17.07/10.58 & 0.24/0.29 & 13.73/8.52 & 17.40/10.82 & 0.28/0.22 & 18.92/13.59 & 23.43/16.25 & 0.22/0.27\\
\hline
ResNet18& P & 11.42/6.96 & 15.18/9.42 & 0.47/0.44 & 13.70/8.49 & 17.32/10.81 & 0.26/0.22 & 18.39/13.60 & 23.86/16.39 & 0.25/0.08\\
\hline
ResNet50& P & 10.30/6.37 & 14.14/8.85 & 0.55/0.51 & 13.77/8.49 & 17.57/10.56 & 0.27/0.22 & 18.51/14.06 & 25.35/16.57 &0.23/0.02 \\
\hline
Inception& P & 10.31/6.24 & 13.86/11.18 & 0.57/0.50 & 13.56/8.29 & 17.42/10.64 & 0.21/0.25 & 18.31/13.53 & 23.24/16.24 & 0.29/0.08\\
\hline
S4 & P & 12.13/7.44 & 14.62/9.96 & 0.44/0.40 & 14.08/8.62 & 17.44/10.46 & 0.24/0.20 & 17.74/12.97 & 24.63/16.55 & 0.30/0.14\\
\hline
Transformer& P & 13.31/6.22 & 14.08/8.70 & 0.55/0.52 & 13.99/8.73 & 17.56/11.11 & 0.22/0.15 & 18.26/13.86 & 23.35/16.66 & 0.23/0.02\\
\hline
MLeNet &P, D& 11.67/7.26 & 14.94/9.31 & 0.37/ 0.40 & 12.43/7.70 & 15.70/9.67 & 0.29/0.33 & 17.12/10.79 & 21.76/13.10 & 0.45/0.38\\
\hline
MResNet18&P, D& 9.41/5.74 & 12.62/7.78 & 0.55/0.58 & 12.11/7.53 & 15.50/9.49 & 0.31/0.35 & 16.94/10.63 & 22.18/12.61 & 0.47/0.44\\
\hline
MResNet50&P, D& 9.13/5.47 & 12.45/7.55 & 0.56/0.60 & 12.39/7.63 & 15.86/9.69 & 0.28/0.33 & 16.89/10.11 & 21.55/12.52 & 0.46/0.49\\
\hline
MInception&P, D& 8.66/5.21 & 11.61/7.14 & 0.62/0.65 & 12.30/7.57 & 15.70/9.61 & 0.30/0.35 & 17.39/10.47 & 21.49/12.90 & 0.45/0.45\\
\hline
MS4 &P, D& 11.48/7.21 & 14.01/8.04 & 0.39/0.42 & 12.50/7.70 & 15.77/9.65 &  0.27/0.32 & 15.40/9.29 & 20.68/12.18 & 0.54/0.54\\
\hline
Ours &P, D& 4.56/2.62 & 6.86/4.10 & 0.87/0.88 & 12.17/7.89 & 15.83/10.06 & 0.24/0.28 & 17.16/11.31 & 20.21/14.12 & 0.43/0.32\\
\hline
\end{tabular}

\label{tab:model_results}
\end{table*}

\subsection{Dataset and Data Pre-processing}
In this work, we utilize the Large scale PulseDB dataset \cite{pulsedb_paper}, which contains 5,245,454 10-second segments of ECG, PPG, and ABP waveforms collected from 5,361 subjects, covering a wide range of physiological and clinical conditions.

To reflect realistic deployment scenarios, we evaluate under three complementary testing protocols following the predefined settings of the PulseDB dataset.
\textit{Calibration-based} testing allows subject overlap between training and testing sets, enabling the model to leverage subject-specific signal characteristics. Performance in this setting reflects generalization to unseen samples from previously observed subjects.
\textit{Calibration-free} testing enforces a strict subject-independent split, where no subjects are shared between training and testing. This setting evaluates the model’s ability to generalize to entirely unseen individuals without subject-specific adaptation.
\textit{AAMI subset} follows the evaluation spirit of the AAMI/ANSI/ISO standard, applying a stricter calibration-free protocol with constrained cohort composition and increased emphasis on blood-pressure distribution tails, providing a rigorous benchmark for medical-device–style assessment.

Prior to model training and evaluation, we apply data cleaning by removing segments containing NaN or infinite values. The Train subset is reduced from 902,160 to 465,480 segments, the Calibration-Based subset from 100,240 to 51,720 segments, the Calibration-Free subset from 111,600 to 57,600 segments, and the AAMI subset from 1,340 to 666 segments.

In addition to temporal segmentation, each PPG wavecycle is assigned a pulse morphology label to characterize whether it exhibits a normotensive-like or hypertensive-like pattern. For each wavecycle, we compute the average values of three morphology-related indices: the augmentation index (AI), the b/a ratio (BA), and the normalized dicrotic notch depth (ND), together with an aggregated morphology score based on previous work \cite{mejia2021photoplethysmography, elgendi2012analysis, zanelli2023clustered}. Specifically, AI quantifies the relative contribution of the reflected wave to the systolic peak, BA reflects waveform steepness, and ND measures the prominence of the dicrotic notch. Based on experimental observations and prior physiological knowledge \cite{simek2005second, otsuka2006utility}, a segment is classified as morphologically normal if the segment-level averages of AI, BA, ND, and morphology score satisfy: AI $\geq 0.5$ and at least two of the following: BA $\geq 1.38$, ND $\leq 0.9$, and morphology score $\geq 0.9$.

Before being fed into the network, each PPG segment is normalized using z-score, enforcing a consistent amplitude scale across subjects and recording sessions and reducing inter-subject variability inherent to optical measurements. Age and BMI are standardized to zero mean and unit variance, while sex is encoded as a binary variable.

\subsection{Experiment settings}
Our network was implemented using the PyTorch and trained on a cluster with four NVIDIA A100 GPUs. We trained models using the Adam optimizer with batch size of 32, betas of (0.9, 0.999), weight decay of $1 \times 10^{-8}$, and fixed learning rate of $2 \times 10^{-5}$. Each model was trained for 100 epochs.

Each PPG segment has a fixed length of 1250 samples. The signal is partitioned into non-overlapping 1D patches using a convolutional patch embedding with a patch size $P=10$ and stride $10$. Each patch is projected into a $C=128$ dimensional embedding space.
The transformer encoder consists of $D=6$ layers. Each layer employs multi-head self-attention with 8 attention heads and an MLP subnetwork with an expansion ratio of 4.
We train two independent models, one for SBP and one for DBP; each model is trained in a multi-task manner, jointly performing BP regression and morphology classification. This design avoids interference between SBP and DBP learning while allowing morphology supervision to regularize feature extraction, thereby simplifying optimization and improving task-specific discrimination, consistent with prior studies \cite{pulsedb_paper, ESMAELPOOR2020103719}. For fair comparison, all baselines were retrained and evaluated under the same one-model-per-BP-component protocol using the same filtered dataset.

\subsection{Results}

\begin{table*}[htbp]
\caption{Ablation study.}
\centering
\begin{tabular}{|l|c|c|c|c|c|c|}
\hline
\multirow{2}{*}{\textbf{Model}}
& \multicolumn{3}{c|}{\textbf{Cal-base (SBP/DBP)}} 
& \multicolumn{3}{c|}{\textbf{Cal-free (SBP/DBP)}}\\
\cline{2-7}
& \textbf{MAE} & \textbf{std} & \textbf{R²} & \textbf{MAE} & \textbf{std} & \textbf{R²} \\
\hline
base &  13.31/6.22 & 14.08/8.70 & 0.55/0.52 & 13.99/8.73 & 17.56/11.11 & 0.22/0.15\\
\hline
b+demo(cat)&  11.42/6.83 & 14.72/ 8.81 & 0.38/0.44 & 12.57/7.93 & 15.98/10.10 & 0.26/0.27\\
\hline
b+demo(FiLM) &  4.73/2.69 & 7.11/4.16 & 0.85/0.87 & 13.28/8.46 & 17.22/10.88 & 0.20/0.20 \\ 
\hline
full &   4.56/2.62 & 6.86/4.10 & 0.87/0.88 & 12.17/7.89 & 15.83/10.06 & 0.24/0.28 \\
\hline
\end{tabular}

\label{tab:ablation}

\end{table*}

Table~\ref{tab:model_results} summarizes the performance of all models under three evaluation protocols: calibration-based (Cal-base), calibration-free (Cal-free), and the AAMI-oriented setting. 

\subsubsection{Calibration-based Performance}

When demographic information is available, our demographic conditioning, morphology-enhanced Transformer achieves the best calibration-based performance across all metrics. In the Cal-base setting, our model reaches a mean absolute error (MAE) of 4.56/2.62 mmHg for SBP/DBP and an $R^2$ of 0.87/0.88. Compared with the strongest demographic-aware baseline: MInception, 8.66/5.21 mmHg and $R^2$ of 0.62/0.65, this corresponds to an approximate 47\% and 50\% improvement in SBP and DBP MAE, respectively, together with a relative $R^2$ improvement of 40\% and 35\%
Relative to representative PPG-only backbones, the improvement is even larger; For example, the PPG-only ResNet50 baseline achieves errors of 10.30/6.37 mmHg with corresponding $R^2$ values of 0.55/0.51. Under the same calibrated protocol, our method improves MAE by approximately 56\% (SBP) and 59\% (DBP), while improving $R^2$ by 58\% and 45\%, respectively. These gains indicate that our proposed demographic-aware attention mechanism, together with morphology-aware supervision, better captures inter-individual variability than both demographic-augmented and PPG-only architectures.

\begin{figure}[t]
    \centering
    \begin{subfigure}[t]{0.241\textwidth}
        \centering
        \includegraphics[width=\textwidth]{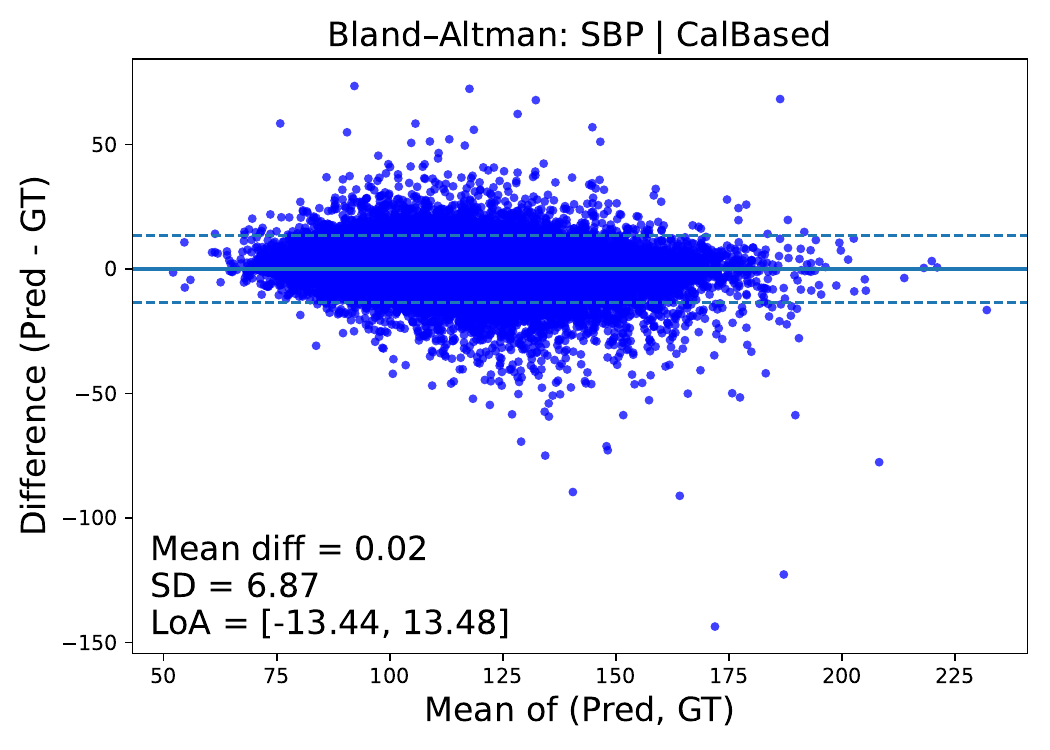}
    \end{subfigure}
    \hfill
    \begin{subfigure}[t]{0.241\textwidth}
        \centering
        \includegraphics[width=\textwidth]{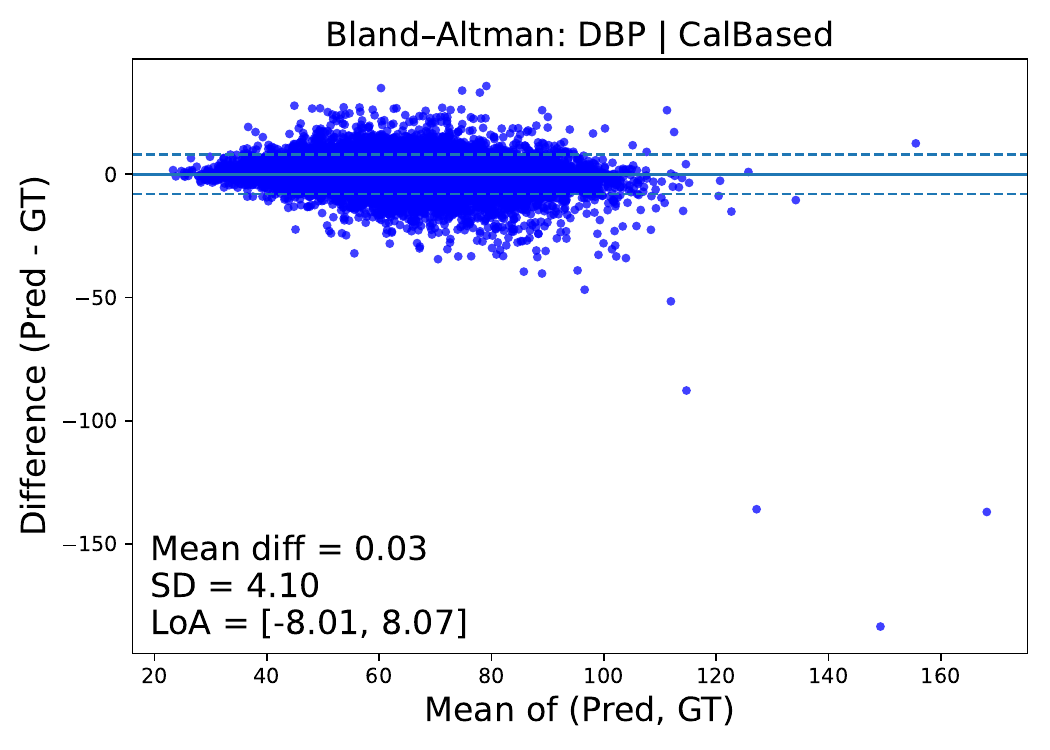}
    \end{subfigure}
    \caption{Bland–Altman analysis of BP estimation performance on Cal-base protocol, with SBP (left) and DBP (right).}
    \label{fig:ba}
\end{figure}

As illustrated in Fig. \ref{fig:ba}, Bland–Altman plots demonstrate that the proposed model exhibits negligible systematic bias and tight limits of agreement for both SBP and DBP in the calibration-based setting.
For SBP, the mean difference between prediction and ground truth is 0.02 mmHg, yielding 95\% limits of agreement of [-13.44, 13.48] mmHg.
For DBP, the mean difference is 0.03 mmHg, with a smaller corresponding limits of agreement of [-8.01, 8.07] mmHg.
In both cases, the error distributions remain approximately centered around zero across the full measurement range, with no evident heteroscedasticity, indicating robust agreement with reference measurements and stable calibration behavior across subjects.

\subsubsection{Calibration-free and AAMI Performance}

In the more challenging calibration-free protocol, where no subject-specific reference is available, the performance gap between models narrows. This may indicate that a substantial fraction of error is driven by subject- and session-specific offsets. Even without calibration, our model remains competitive: it achieves a Cal-free MAE of 12.17/7.89 mmHg with $R^2$ of 0.24/0.28, matching the overall level of strong demographic-aware baselines. A similar pattern appears under the AAMI-like evaluation, where we obtain 17.16/11.31 mmHg with $R^2$ of 0.43/0.32. Taken together, these results position our conditional Transformer firmly in the leading group, especially for practical deployments where brief onboarding calibration is acceptable, while also motivating targeted robustness enhancements to further strengthen cross-subject generalization.

\subsection{Ablation Study}

Table~\ref{tab:ablation} summarizes four models designed to isolate the effect of each part of our proposed mode. The \textit{base} model is a vanilla Transformer that receives only PPG signal. \textit{b+demo(cat)} augments this baseline by concatenating demographic features to the PPG signal feature before the attention mechanism. \textit{b+demo(FiLM)} replaces concatenation with FiLM-based conditioning, where demographics generate layer-wise modulation parameters injected into all attention layers. Finally, \textit{full} corresponds to our complete framework.

Starting from \textit{base}, we observe large errors and weaker $R^2$ in both cal-based and cal-free settings, highlighting the difficulty of capturing inter-subject variability from PPG alone. Introducing demographics via concatenation (\textit{b+demo(cat)}) brings modest improvements, especially in the cal-free regime where SBP/DBP MAE and $R^2$ both improve over \textit{base}. 
In contrast, FiLM-based conditioning (\textit{b+demo(FiLM)}) yields a pronounced benefit in the cal-based setting: SBP and DBP MAE drop to around 4.7/2.7\,mmHg, accompanied by lower std and substantially higher $R^2$. This indicates that demographic priors are effective when they modulate intermediate representations throughout the network, guiding attention and feed-forward layers to adapt to demographic patterns. In the cal-free setting, \textit{b+demo(FiLM)} still has a slight improvement on the baseline, reflecting that the method can decrease the challenge of generalizing demographic-conditioned modulation.
The \textit{full} model achieves the best overall performance by adding auxiliary morphology head with learnable task weights. Overall, the ablation studies demonstrate that both FiLM-based demographic modulation and morphology supervision make meaningful contributions, and their integration produces the most robust and physiologically informed BP estimator.

\section{Discussion}

Our model achieves a substantial gain in the calibration-based protocol, which closely matches realistic deployment workflows. In many care pathways, a single cuff measurement can be obtained with minimal burden during clinic intake, discharge, or device onboarding. After which continuous cuffless monitoring can provide high-frequency tracking of BP dynamics in daily life \cite{Derendinger2024Track, Sola2022Guidance}. 
This shift from sporadic point measurements to dense longitudinal trajectories is clinically valuable: it can reveal sustained elevation, increased variability, or emerging deviations earlier than intermittent cuffs, thereby enabling timelier follow-up, medication adjustment, or lifestyle interventions. Importantly, the proposed approach provides the groundwork for lightweight and single-sensor deployment, supporting scalable integration into wearable devices and smartphones through real-time edge inference on accessible PPG signals with minimal hardware overhead.

Yet improvements are less consistent under Cal-free and AAMI tests. Calibration can absorb subject- and device-specific biases, whereas Cal-free/AAMI settings require invariant cues that transfer to unseen subjects and conditions. This challenge is amplified by noisy PPG and a segment-heavy dataset: only thousands of subjects have complete demographics versus hundreds of thousands of segments, which magnifies subject/device/context shifts. AAMI criteria further stress-test reliability by emphasizing error distribution and consistency, making them sensitive to long-tail outliers. Future work should focus on subject-diverse curation, subject-balanced sampling, and robust/adaptive training to achieve better generalizability on Cal-free/AAMI protocols.

\section{Conclusion}

In this work, we presented a demographic conditioning, morphology-enhanced Transformer for cuffless BP estimation from PPG signal. By combining a 1D convolutional patch embedding stem with FiLM-based demographic conditioning in every Transformer layer, our model jointly leverages long-range waveform dynamics and demographic information. An auxiliary morphology head further regularizes the backbone toward hypertension-related pulse-shape patterns, reducing reliance on amplitude-only cues and improving physiological interpretability.
On the PulseDB dataset, the proposed model achieves substantial gains over baselines, particularly in the calibration-based protocol. It attains 4.56/2.62~mmHg MAE and $R^2$ of 0.87/0.88 for SBP/DBP, corresponding to about 50\% MAE reduction relative to the best demographic-aware network. 
Overall, our framework enables scalable, single-sensor BP monitoring for longitudinal trend tracking and early risk signaling in remote and ambulatory settings, with only a low-burden calibration during onboarding.

\section*{Acknowledgment}
This work is partially supported by NSF grants CCF-2504152,  CNS-2107057 and CTSA grant UM1TR004539. 
This study used PulseDB, a publicly available de-identified dataset derived from MIMIC-III and VitalDB. No new participants were recruited, no new physiological recordings were collected, and no animal experiments were conducted. This work is a secondary analysis of publicly available de-identified data and did not require additional informed consent.

\bibliographystyle{ieeetr}  
\bibliography{example}

\end{document}